\documentclass[aps,prl,twocolumn,showpacs,superscriptaddress, twocolumn, amssymb,showpacs,nofootinbib,tighten]{revtex4}
\usepackage{amssymb,amsmath,graphicx,color,microtype}
\usepackage{graphicx}
\usepackage{txfonts}

\usepackage{epsf}
\usepackage{epstopdf}
\usepackage {amssymb}
\newcommand{\nc}{\newcommand}
\nc{\ba}{\begin{eqnarray}}
\nc{\ea}{\end{eqnarray}}
\newcommand\be{\begin{equation}}
\newcommand\ee{\end{equation}}

\nc{\e}{{\bf{e}}}
\nc{\kk}{{\bf{k}}}
\nc{\pp}{{\bf{p}}}

\nc{\bfk}{{\bf{k}}}
\nc{\bfx}{{\bf{x}}}
\nc{\bfp}{{\bf{p}}}

\nc{\eH}{{\epsilon_H}}
\nc{\calP}{{\cal P}}
\nc{\im}{{ \mathrm{Im} } }

\begin{document}
\title{Evidence of Neutrino Enhanced Clustering in a Complete Sample of
Sloan Survey Clusters, Implying $\sum m_{\nu}= 0.119 \pm 0.034$ eV }

\author{Razieh~Emami}
\email{razieh.emami$_{}$meibody@cfa.harvard.edu}
\affiliation{Center for Astrophysics, Harvard-Smithsonian, 60 Garden Street, Cambridge, MA 02138, USA}
\affiliation{ Institute for Advanced Study, The Hong Kong University of Science and Technology, Clear Water Bay, Kowloon, Hong Kong}

\author{Tom~Broadhurst}
\affiliation{Department of Theoretical Physics, University of the Basque Country UPV-EHU, 48040 Bilbao, Spain}
\affiliation{IKERBASQUE, Basque Foundation for Science, Alameda Urquijo, 36-5 48008 Bilbao, Spain}

\author{Pablo~Jimeno}
\affiliation{Department of Theoretical Physics, University of the Basque Country UPV-EHU, 48040 Bilbao, Spain}

\author{George~ Smoot}
\affiliation{ Helmut and Anna Pao Sohmen Professor-at-Large, IAS, Hong Kong University of Science and Technology,
Clear Water Bay, Kowloon, 999077 Hong Kong, China}
\affiliation{Paris Centre for Cosmological Physics, APC, AstroParticule et Cosmologie, Universit\'{e} Paris Diderot,
CNRS/IN2P3, CEA/lrfu, Observatoire de Paris, Universit\'{e} Sorbonne Paris Cit\'{e}, 10, rue Alice Domon et Leonie Duquet,
75205 Paris CEDEX 13, France}
\affiliation{Physics Department and Lawrence Berkeley National Laboratory, University of California, Berkeley,
94720 CA, USA }

\author{Raul~Angulo}
\affiliation{Centro de Estudios de F ́ısica del Cosmos de Arag ́on, Plaza de San Juan 1, 44001 Terue, spain}

\author{Jeremy~Lim}
\affiliation{Department of Physics, The University of Hong Kong, Pokfulam Road, Hong Kong}
\affiliation{Laboratory for Space Research, Faculty of Science, The University of Hong Kong, Pokfulam Road, Hong Kon}

\author{Ming Chung~Chu}
\affiliation{Department of Physics, The Chinese University of Hong Kong,
Shatin, N.T., Hong Kong, China}

\author{Shek~Yeung}
\affiliation{Department of Physics, The Chinese University of Hong Kong,
Shatin, N.T., Hong Kong, China}

\author{Zhichao Zeng}
\affiliation{Department of Physics, The Chinese University of Hong Kong,
Shatin, N.T., Hong Kong, China}
\affiliation{Department of Physics, Ohio State University, Columbus, Ohio 43210, USA}

\author{Ruth~Lazkoz} 
\affiliation{Department of Theoretical Physics, University of the Basque Country UPV-EHU, 48040 Bilbao, Spain}

\begin{abstract}

The clustering amplitude of 7143 clusters from the Sloan Digital Sky Survey (SDSS) is found to increase with cluster mass, closely agreeing with the Gaussian random field hypothesis for structure formation. The amplitude of the observed cluster correlation exceeds the predictions from pure cold dark matter (CDM) simulation by $\simeq 6\%$ for the standard Planck-based values of the cosmological parameters. We show that this excess can be naturally accounted for by free streaming of light neutrinos, which opposes gravitational growth, so clusters formed at fixed mass are fewer and hence more biased than for a pure CDM density field. An enhancement of the cluster bias by 7\% matches the observations, corresponding to a total neutrino mass, $m_{\nu} = 0.119 \pm 0.034$ eV at 67\% confidence level, for the standard relic neutrino density. If ongoing laboratory experiments favor a normal neutrino mass hierarchy then we may infer a somewhat larger total mass than the minimum oscillation based value, $\sum m_{\nu} \simeq 0.056eV$, with 90\% confidence. Much higher precision can be achieved by applying our method to a larger sample of more distant clusters with weak lensing derived masses. 
\end{abstract}

\maketitle

\section{Introduction} 

The standard picture whereby cosmic structure develops gravitationally from a Gaussian random field (GRF) predicts that the amplitude of clustering should increase steadily with density contrast \cite{Kaiser:1984sw, Bardeen:1985tr, Cole:1989vx, Mo:1995cs, Sheth:1999mn}. This inherent property of a GRF is known as the clustering ``bias", and was first employed by \cite{Kaiser:1984sw} to explain the approximately factor of three larger correlation length of nearby massive galaxy clusters relative to field galaxies. Subsequent N-body simulations have fully demonstrated that collapsed halos formed from a GRF should be biased in this fundamental way \cite{ Dalal:2008zd}. Simulations also predict that the clustering bias should increase with redshift as halo abundance declines with redshift, particularly for the most massive clusters. The effect of light neutrinos is to smooth the density field thereby slowing the growth of structure, thus reducing the abundance of clusters \cite{Eisenstein:1998tu} while enhancing the clustering amplitude relative to a pure CDM density field.

Measuring this fundamental link between halo mass and clustering bias has not proven feasible using galaxies.  Not only are the virial masses of galaxies hard to define observationally, but galaxy correlation functions have been found to depend strongly on galaxy type and luminosity with a complexity that cannot be simply linked to the growth of structure. In contrast, cluster masses can be directly inferred from gravitational lensing and correlate almost linearly with the number of member galaxies, following a clear mass-richness (MR) relation \cite{Jimeno:2017ecv,Rozo:2011xj, Baxter:2016jdq, Rykoff:2016trm, Geach:2017crt}.

Large surveys are now underway to realize the anticipated sensitivity of cluster abundance to the growth of the cluster mass function \cite{Bahcall:2003hu}, including a predicted small additional suppression of their numbers by neutrino free-streaming \cite{Banerjee:2016zaa, Costanzi:2013bha, Raccanelli:2017kht}. This suppression is claimed to be close to detection in an initial SZ-selected sample of 370 clusters \cite{deHaan:2016qvy} and in combination with other methods provides $\sum m_{\nu}\lesssim 0.14ev$ (\cite{Vagnozzi:2017ovm, Alam:2016hwk}), tightening the robust 95\% upper limit of $<0.25$ eV from the pure Planck analysis \cite{Ade:2013zuv}. This claimed improvement over Planck rests on uncertain assumptions about the inherent spread in mass of strong SZ detected clusters, as cluster collisions compress the gas thereby boosting the SZ signal \cite{Molnar:2014rqa, Molnar:2017man}\cite{Menanteau:2010zi}.

Here we develop and apply a clustering based method that is sensitive to the effect of neutrinos, using the correlation length of optically detected clusters from the thoroughly tested and currently largest and most complete survey of SDSS clusters identified by the RedMapper team \cite{Rykoff:2013ovv, Rozo:2013vja, Rozo:2014zva}. This large RedMapper has made redshift complete in our earlier work \cite{Jimeno:2016nbf, Jimeno:2017ecv} by cross correlation with the BOSS spectroscopic data, which includes over 7000 clusters for which precise correlation functions have been estimated  \cite{Jimeno:2016nbf, Jimeno:2017ecv}. The correlation function scales as the square of the bias and, as we show here, is already sensitive enough to detect the effects of neutrino mass inferred from the oscillation experiments of $\sum m_{\nu} \simeq 0.056 eV$.

We rely on the latest Planck determinations of all the cosmological parameters \cite{Ade:2015xua} required here for the theoretical calculation of the bias enhancement by neutrinos and for the comparison with the MXXL cosmological simulations, where for a flat cosmology,  $\Omega_{\Lambda} = 0.6911$, $\Omega_m = 0.3089$, $\Omega_b = 0.0497$, $h = 0.6773$ and the amplitude of the primordial power-spectrum $A_s = 2.142 \times 10^{-9}$, and hence the only flexible quantity is the cold dark matter density, defined as, $\Omega_c = \Omega_m - \Omega_b - \Omega_{\nu}$, with $\Omega_{\nu} h^2\equiv \frac{ (1.015)^{3/4} m_{\nu}}{94.07~ev} $. We havIn this analysis we examine first one species of dominantly massive neutrino, corresponding to the standard neutrino hierarchy, and then we discuss implications for the inverted hierarchy, which may be favored by our results. 
We emphasis here that we are safe from the simulation to simulation scatter discussed by Tinker \cite{Tinker:2010my} owing to the relatively narrow halo mass and redshift range that we consider. 
\section{Neutrino Bias Enhancement for Clusters}
We first present a consistent formalism for the biasing effect of light neutrinos on cluster scales, emphasizing that the cluster correlation provides a relatively sensitive way of detecting the effect of standard relic neutrinos. We then compare the cluster correlation function predicted by our analysis with that measured for the galaxy clusters from the SDSS selected in the manner described above.

\subsection{Theoretical Considerations}

Here we estimate the effect of light neutrinos on the correlation function of clusters using 
the empirically peak background split approximation \cite{Sheth:1999mn}, hereafter Sheth-Tormen (ST). \\

%
In the ST approach, the bias is defined as,
\ba
\label{ST-bias1}
b_{ST} \equiv \frac{a \nu_1 -1}{\delta_c} + \frac{2p/\delta_c}{1 + (a \nu_1)^p},
\ea
here $\nu_1 \equiv \left(\delta_c/\sigma_0\right)^2$ where $\delta_c$ refers to the critical over-density. In addition, $\sigma_0^2 \equiv  \left(\frac{1}{2 \pi^2}\right) \int dk k^{2} P(k) W^2(kR)$ denotes the 0-th spectral moment of the matter power-spectrum and we use the top hat filter function, $W(kR) = 3\left(\sin{(kR)} - kR\cos{(kR)}\right)/(kR)^3$, as a smoothing factor. We use the usual simulation calibrated free parameter preferences, $a = 0.707, p = 0.3$. 
We compute the 0-th spectral moment of the matter power-spectrum as well as the critical over-density. 
The former can be calculated using the matter power spectrum, conveniently given by the CAMB code \cite{Lewis:2002nc, Lewis:2013hha}. 

To estimate the critical over-density for cluster formation, we determine the linearly evolved value of the initial over-destiny required to have a collapse at red-shift $z$. Adopting spherical collapse, which is a good estimate for heavy halos, we then calculate the evolution of the CDM + Baryon over-density, $ \delta_{cb} \equiv \left(\bar{\rho}_c \delta_c + \bar{\rho}_b \delta_b\right)/\left(\bar{\rho}_c + \bar{\rho}_b\right)
$ as, 
\ba
\ddot{\delta}_{cb} + 2 H \dot{\delta}_{cb} - \frac{4}{3} \frac{\dot{\delta}^2_{cb}}{1 +\delta_{cb} } - \frac{3H^2_0 }{2} \left( \Omega_{cb} \delta_{cb} + \Omega_{\nu} \delta_{\nu}\right) \left(1 + \delta_{cb}\right) &=& 0, \nonumber\\
\ddot{R} + \frac{G M}{R^2} + H^2_0\left( \Omega_r a^{-4} + \frac{\rho_{\nu}p_{\nu}}{3H^2_0 M^2_{P}} - \Omega_{\Lambda}
\right) R + \frac{G \delta M_{\nu}(<R)}{R^2}&=& 0, \nonumber\\
\ea
where $\rho_{\nu} = 2 \int \frac{d^3 p }{(2\pi)^3} \frac{\sqrt{p^2 + m^2_{\nu}}}{\exp{(p/T_{\nu})}+1}$ refers to the neutrino energy density and 
$p_{\nu} = 2 \int \frac{d^3 p }{(2\pi)^3} \frac{p^2}{\sqrt{p^2 + m^2_{\nu}}}\frac{1}{\exp{(p/T_{\nu})}+1}$ denotes the neutrino pressure. Here $T_{\nu} = \left(\frac{1.95491}{a}\right)~K$ is the physical temperature.  
In addition, $\Omega_{cb}\equiv\left(\Omega_{c} + \Omega_{b}\right)a^{-3}, ~\delta_{\nu}\equiv\delta M_{\nu}(<R) / \left(4 \pi R^3 H^2_0 M^2_{P} \Omega_{\nu}\right)$ in which $\delta M_{\nu}(<R) =  m_{\nu} \int_{V_c} d^3 \mathbf{r} \int \frac{d^3 \mathbf{q}}{(2\pi)^3} f_1(\mathbf{q}, \mathbf{r}, t)$
refers to the neutrino mass function interior to the radius $R$. $f_1$ is the linear perturbation to the neutrino distribution function, $f = f_0 + f_1$, which can be calculated from the evolution of the radius $R$, in the absence of the neutrino clustering, \cite{LoVerde:2014rxa}, as, 
\ba
&& f_1 (\mathbf{r}, \mathbf{q}, \eta ) = 2 \frac{m_{\nu}}{T_{\nu}} \int_{t_0}^t \frac{dt'}{a(t')} \frac{\exp{(q/T_{\nu})}}{den\left(1 + \exp{(q/T_{\nu})} \right)^2}
\frac{G \delta M(t')}{r^2} \nonumber\\
&& \times \left(\alpha \frac{q}{T_{\nu}} - \hat{q}\cdot \hat{r}\right) \bigg{[}
\frac{a^3(t') r^3}{R^3} \Theta \bigg{(} r^2 \left( 1 + q^2/T^2_{\nu} \alpha^2 - 2q/T_{\nu} \alpha
\right) < \nonumber\\
&& \frac{R^2(t')}{a^2(t')}\bigg{)}
+ \frac{\Theta \left( r^2 \left( 1 + q^2/T^2_{\nu} \alpha^2 - 2q/T_{\nu} \alpha \hat{q}\cdot \hat{r}
	\right) \geq \frac{R^2(t')}{a^2(t')}\right) }{\left(  1 + q^2/T^2_{\nu} \alpha^2 - 2q/T_{\nu} \alpha \hat{q}\cdot \hat{r}  \right)^{3/2}}
\bigg{]}. \nonumber\\
\ea
where $\delta M(t) = M - \frac{4 \pi}{3}\bar{\rho}_{cb} R^3$ and $\Theta$ denotes the Heaviside step function. In addition, $\alpha$ is defined as $\alpha \equiv T_{\nu} \left(\eta -\eta'\right) / (m_{\nu} r)$ and $\eta$ is defined as, $a^2d\eta = dt$, note that both $r$ and $q$ are comoving quantities. 
These are the set of the differential equations that must be solved together with the evolution of the Hubble parameter: 
$
H^2 = H_0^2 \bigg{(} \Omega_{\gamma} a^{-4} + (\Omega_{c}+ \Omega_b) a^{-3}  + \Omega_{\Lambda} +
\frac{\rho_{\nu}}{{(3 M^2_{P} H_0^2)}} 
\bigg{)}
$

We now describe the initial conditions, for which there are three independent, relevant quantities, $\delta_{cb, ini}, \dot{\delta}_{cb, ini}, a_{ini}$. The first determines the collapse redshift that relates to second condition via: $\dot{\delta}_{cb,ini} = \delta_{cb,ini} \left( \frac{d \ln{\delta_{cb}}(t)}{dt}\right){|}_{,ini}$, where the term in the parenthesis is free of any normalization and can be extracted from the CAMB code. Initiating the evolution at $z \simeq 200$, we also read off the value of the initial scale factor. Using the above quantities, we can obtain $R_{ini} = \bar{R}_{ini} \left(1 - \frac{1}{3}\delta _{cb,ini}\right)$ and $\dot{R}_{ini} = H_{ini} \bar{R}_{ini} \left(1 - \frac{1}{3}\delta _{cb,ini} -\frac{1}{3} H^{-1}_{ini} \dot{\delta}_{cb,ini} \right)$, where $\bar{R}_{ini}  = \left(\frac{3 M}{4 \pi \rho_{cb,ini}}\right)^{1/3} $ depends on the value of the halo mass. 

Next, we compute the evolution of system. The only unknown quantity is the redshift of collapse. We treat this as a perturbation to an Einstein-de (Ed) Sitter universe with a constant critical over-density \cite{Diemer:2017bwl, Gunn:1972sv}, $\delta_{c,Ed} = 1.68647$, by truncating the evolution at a redshift once the non-linear over-density for any cosmology reaches this number. We then read off the linearly evolved over-density at this reference redshift as the critical over-density. We checked the robustness of our critical over-density by choosing another method in which we write the spherical collapse equations in terms of $1/\delta$ instead of $\delta$, thus removing the need to specify the actual collapse. Comparing the results of these two methods, we found a difference of only 0.01\% thus ensuring that the results are independent of the method we use. We now have the information necessary to compute the bias for ST. 

\begin{figure}[!h]
\centering
\includegraphics[width=0.5\textwidth]{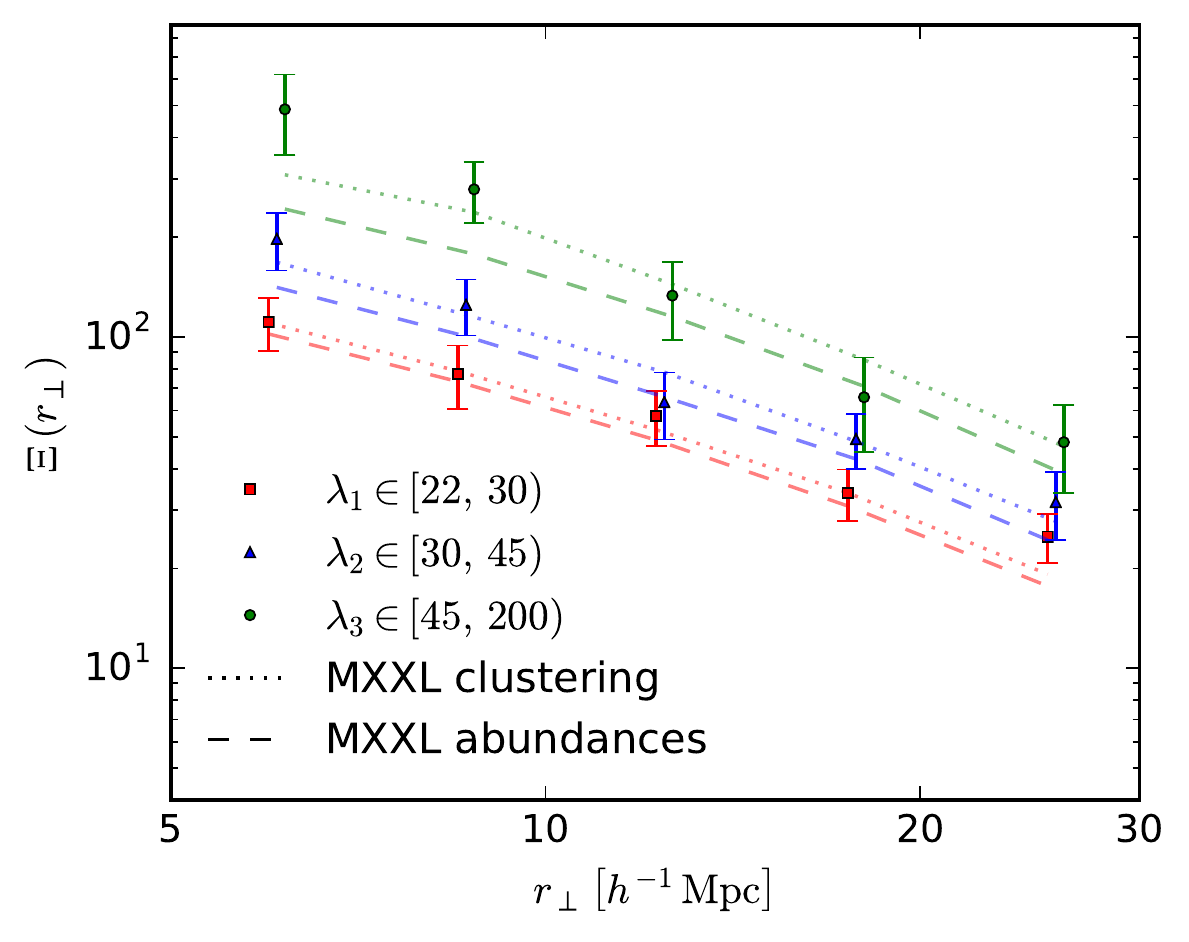}
\caption{Measured correlation functions of ReDMaPPer clusters for 3 richness ranges, showing clearly the observed amplitude scales with mass at almost the same rate as the predicted by the MXXL simulation (dotted and dashed curves) confirming directly the fundamental Gaussian field hypothesis for the formation of structure. In detail, when averaged, the data lies significantly higher than the pure CDM based prediction based on a fit to the cluster mass function from MXXL, by $\simeq 6\% $ (dashed curves). }
	\label{combinedAbun}
\end{figure}

\subsection{N-body Simulation}

Here we check our analytical predictions against our new cosmological simulations in the presence of massive neutrinos, following the grid-based method described in \cite{Zeng:2018pcv, AliHaimoud:2012vj}. We combine the GADGET2 N-body code \cite{Springel:2005mi} for CDM particles and simulate relic neutrinos as a grid-based density field, where only the long-range force (the PM part) is re-calculated by incorporating the effects of neutrinos: $\delta_{tot} = (1-f_{\nu}) \delta_{c+b} + f_{\nu} \delta_{\nu}$, where $f_{\nu}$ is the mass fraction of cosmological neutrinos, and $\delta_{tot}$, $\delta_{c+b}$ and $\delta_{\nu}$ are the over-density fields of, respectively, total matter, CDM and baryons, and neutrinos.  $\delta_{\nu}$ follows the linear perturbation theory with the potential provided by all matter, whereas $\delta_{c+b}$ is calculated in GADGET2 and then replaced by $\delta_{tot}$ after we take this weighted average. We iterate to advance $\delta_{\nu}$ and $\delta_{tot}$ using the linear perturbation equation and GADGET2.
We have set up four realizations for each neutrino mass, each realization having a box size of 800 \rm{Mpc/h} and $512^3$ simulation particles. The halo catalogues are generated by Amiga Halo Finder \cite{Knollmann:2009pb}.

\subsection{Comparison with Observations}

 The carefully constructed and well tested sample of SDSS clusters as defined by the RedMapper team \cite{Rykoff:2013ovv, Rozo:2013vja, Rozo:2014zva} has provided many insights, including accurate lensing and SZ based mass-richness relations \cite{Jimeno:2017ecv, Geach:2017crt, Simet:2016mzg}. Redmapper clusters are determined to be complete to $z=0.33$ in terms of detectability within the SDSS above a minimum richness of $\lambda>20$ over this redshift range \cite{Rykoff:2013ovv, Rozo:2013vja, Rozo:2014zva}. This claim is strongly supported by \cite{Jimeno:2016nbf}, where the observed numbers of RedMapper clusters are proportional to the cosmological volume in the redshift range $z<0.33$, but above which their numbers markedly decline relative to the available volume. Importantly, this volume complete sample of Redmapper clusters has been made redshift complete by \cite{Jimeno:2016nbf} for the first time, at a level of 97\% within the above RedMapper limit $z<0.33$, by cross correlation with the latest SDSS/BOSS spectroscopic redshift surveys, allowing correlation functions to be accurately measured on small scales using these accurate redshifts \cite{Jimeno:2016nbf}.

The 2-point cluster correlation function of this complete sample is shown in Figure \ref{combinedAbun}, where it is clear that the amplitude of this correlation function increases with cluster richness. The correlation function has been integrated on small scales along the line of sight to account for the well known velocity effects that otherwise affect pairwise distance separations. We compare this data with the predicted correlation functions from the large, MXXL simulation of pure CDM \cite{Angulo:2012ep}. This simulation is scaled by the latest Planck cosmological parameters, following the prescription of \cite{Angulo:2012ep}, for the same redshift range as the complete RedMapper sample, and integrated along the line of sight in the same way as the data. Two different sets of simulation based predictions are calculated, where the lower curves shown in Figure \ref{combinedAbun} are determined by a joint fit to the abundance of RedMapper clusters as a function of richness whereas the upper curves are fitted to the clustering. For both of the above simulations a standard power-law function for the mass-richness relation is used for this transformation with three free parameters, slope, scatter and the pivot point normalization (see \cite{Jimeno:2016nbf} for details), that are solved for simultaneously in comparing the measurements with the MXXL simulation.

It is clear from the comparison in Figure \ref{combinedAbun} that the observed cluster correlations are close to the MXXL predictions in slope and amplitude and the scaling with richness shown in Figure \ref{combinedAbun} for the 3 independent richness bins, implying excellent agreement with the GRF hypothesis for the origin of cosmic structure. In detail we see the abundance-based prediction (solid curves) is systematically below the data and below the cluster-based fit (dashed curves). This mismatch was highlighted by \cite{Jimeno:2016nbf} without any satisfactory resolution, with the difference between these two mass-richness relations found to be mainly in the normalization, with the correlation-function based predicted cluster being $\simeq 56\%$ higher in mass, $M_{200m}=4.7\times 10^{14}M_{\odot}/h$, than the abundance-based mean cluster mass $(3.02\pm 0.11) \times 10^{14}M_{\odot}/h$. This is a large difference in mass and reflects the relatively shallow dependence of clustering amplitude with cluster mass, so that a significantly higher mean mass is required to match the observed clustering level but corresponds to a much lower cluster abundance as the cluster mass function is inherently steep. Furthermore, the larger mean mass of the correlation function fit (shown in Figure \ref{combinedAbun}) is excluded by the independent weak-lensing based mass-richness relation derived by \cite{Simet:2016mzg} for the same RedMapper  sample, where the mean mass is just $(2.70\pm0.2) \times 10^{14}M_\odot/h$ with a 7\% estimated uncertainty \cite{Simet:2016mzg}, and therefore consistent within $1.3\sigma$ with our aforementioned abundance-based zero-point mass. This WL based mass is also supported by a new mass-richness related analysis performed in  \cite{Murata:2017zdo} who derive a mean mass of $3.0\times 10^{14}M_\odot/h$ using WL measurements for RedMapper clusters. This agreement of independent weak lensing based masses with our abundance-based best fitting mean mass further motivates our exploration of the effect of light neutrinos: since, as we derived in the previous section, an enhancement in bias is expected to be significant in terms of reconciling our observed mass function with the observed  correlation function without the need for increasing cluster masses.


We now show how our excess correlation may be explained by the neutrino induced bias derived above. The relation between the 2-point mass correlation function of collapsed halos relates simply to the bias of a GRF via: $\xi_{HH}(r) \equiv b^2\xi_{MM}(r)$, where $\xi_{HH}$ denotes the correlation of the collapsed halos, and $b$ is the bias. This relation may be linked to the mass power spectrum via $\xi_{MM}(r) \equiv \left(\frac{1}{2 \pi^2}\right) \int dk k^2 P(k) W(kr)$, where $W(kr)$ refers to the top hat filter function. 
Observationally, we estimate the above correlation function with the following power-law scaling: $\xi(r)_{obs}=(r/r_0)^{-\gamma}$, of correlation length $r_o$ and slope of $\gamma=1.7\pm 0.05$. 
\begin{figure*}[ht!]
 \center
 \vspace{-1pt}
  \includegraphics[width=\textwidth]{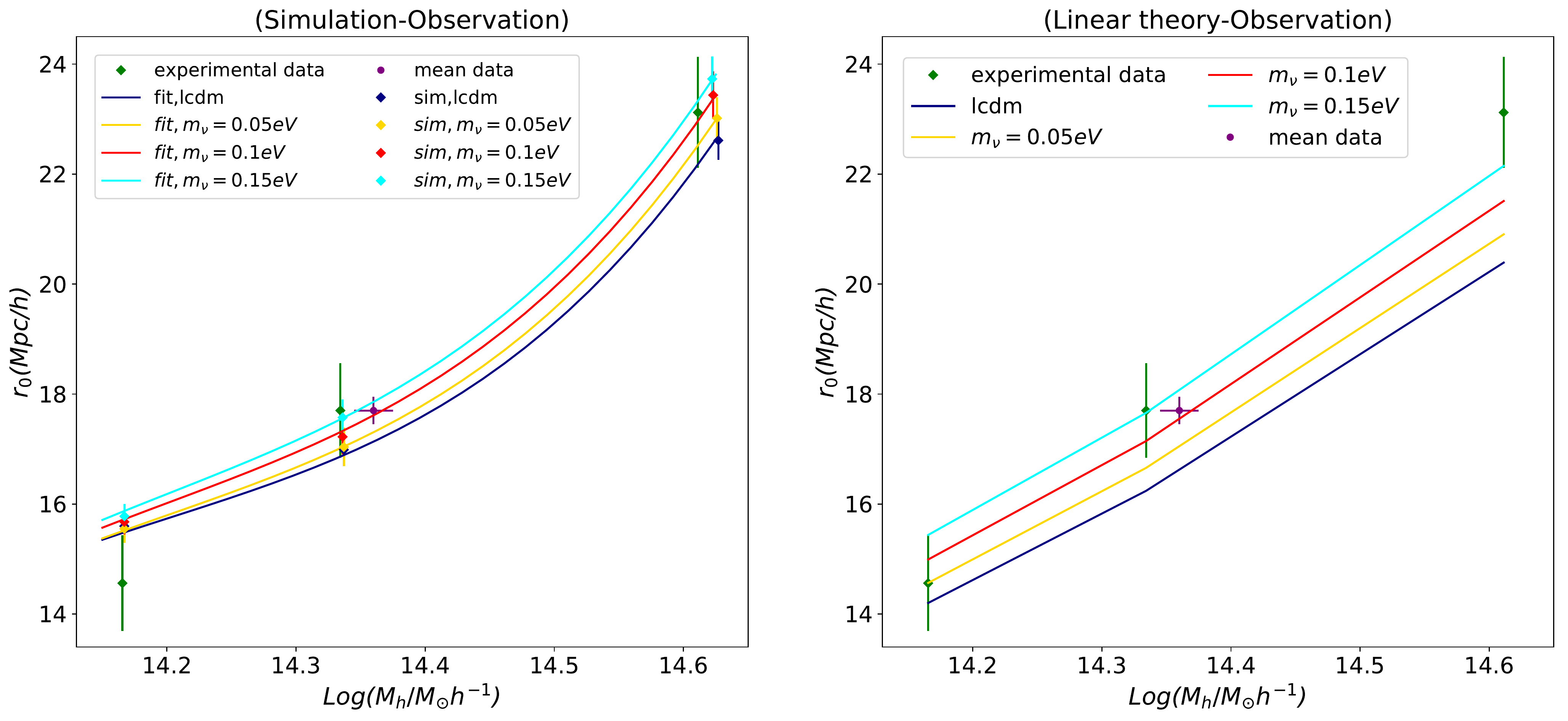}
  \vspace{-4pt}
\caption{\textit{(Left)} comparison between the correlation length $r_{0}$ from the simulation and from 3 independent richness bins (green points) and the data mean (purple point) as a function of halo mass. The simulation points are shown with different colors. \textit{(Right)} comparison between the linear theory and the RedMapper results. The predicted neutrino mass from the linear theory is similar to that from the non-linear simulation, with a somewhat weaker mass dependence. 
}
\label{r0-value}
\end{figure*}

To compare the RedMapper observations with theory, we are required to use the mass-richness relation.  This link is obtained simultaneously in the fits, as described in Ref. \cite{Jimeno:2016nbf} - where we fit the halo mass function from the simulations leaving free the 3 parameters of the MR relation, namely the zero point, the slope and the dispersion, as shown in figure 14 of \cite{Jimeno:2016nbf}. The values of these 3 parameters are in very good agreement with those independently derived using lensing, X-ray, and SZ based MR relations in Refs. \cite{Simet:2016mzg, Jimeno:2017ecv}. These independent derivations were made subsequent to our work, providing verification that adds great confidence to our best-fitting neutrino mass.  

We can make a rough consistency check of this result that is independent of the simulation based comparison used above by looking at the absolute value of  $r_{0}$ from a weak-lensing based mass-richness relation for this same cluster sample derived in \cite{Simet:2016mzg}. In order to calculate this quantity, we use the fact that at $ r = r_{0}$ the halo-halo correlation function is, by definition, unity, and seek $b_{ST}^2\left(\frac{1}{2 \pi^2} \int dk k^2 P(k) W(kr)\right) = 1$ for which we use the ST bias as the input. The results are given in Fig. \ref{r0-value}. In the left panel, we present the relation between the fitted correlation length $r_0$ and the halo mass in these sets of simulation, averaged over four realisations. To be consistent with the RedMapper data, the shown results are also averaged over the redshift range of 0.1 to 0.3. In the right panel, we show the analytical results the $r_0$. Both the analytical and simulation predictions are in close accord.  The impact of the massive neutrinos on enhancing the halo clustering is clearly seen from these plots. The mean observational value (marked in red point) is slightly higher than the pure CDM predictions corresponding to the range  $0.08eV  < \sum m_{\nu} <0.13eV$.
 


\subsection{Discussion and Conclusions}

We have shown here that even the minimum mass density of standard relic neutrinos is expected to enhance the clustering length of galaxy clusters by at least 3\%, relative to pure cold dark matter. This bias induced boost to the clustering of clusters is more than an order of magnitude larger than the effect of neutrinos on the general power spectrum of galaxies because the clustering bias of clusters is approximately three times that for galaxies and the correlation function amplitude scales as the square of this bias. Furthermore, the sign of this boost
in the clustering of halos has the {\it opposite} sign to the usually sought suppression of the matter power spectrum of the mass density field, which is predicted to ``step" down on scales below the predicted free streaming scale of $\lesssim  100 \rm{Mpc}$. This enhancement of the halo correlation function is noted in the simulations of \cite{Marulli:2011he} and also visible in Figure 3 of \cite{Rizzo:2016mdr} when incorporating light neutrinos.

The carefully defined RedMaPPer clusters from the SDSS with full spectroscopic redshifts by \cite{Jimeno:2017ecv} provides a large, complete sample of clusters with accurate redshifts. Using this data we have claimed a 6\% bias enhancement of the correlation length of clusters, corresponding to total neutrino mass of $\sum m_{\nu}=0.119 \pm 0.034 eV$ at 67\% confidence level, for the standard relic mass density and standard Cosmological parameters. This lies well below the robust bound of $<0.24 eV$ from the Planck collaboration (2015), and below subsequent claimed improvements on this upper limit that incorporate additional constraining data from cluster counts \cite{deHaan:2016qvy} and the Lyman-$\alpha$ forest \cite{Seljak:2006bg}.

\begin{figure*}[ht!]
 \center
\includegraphics[width=\textwidth]{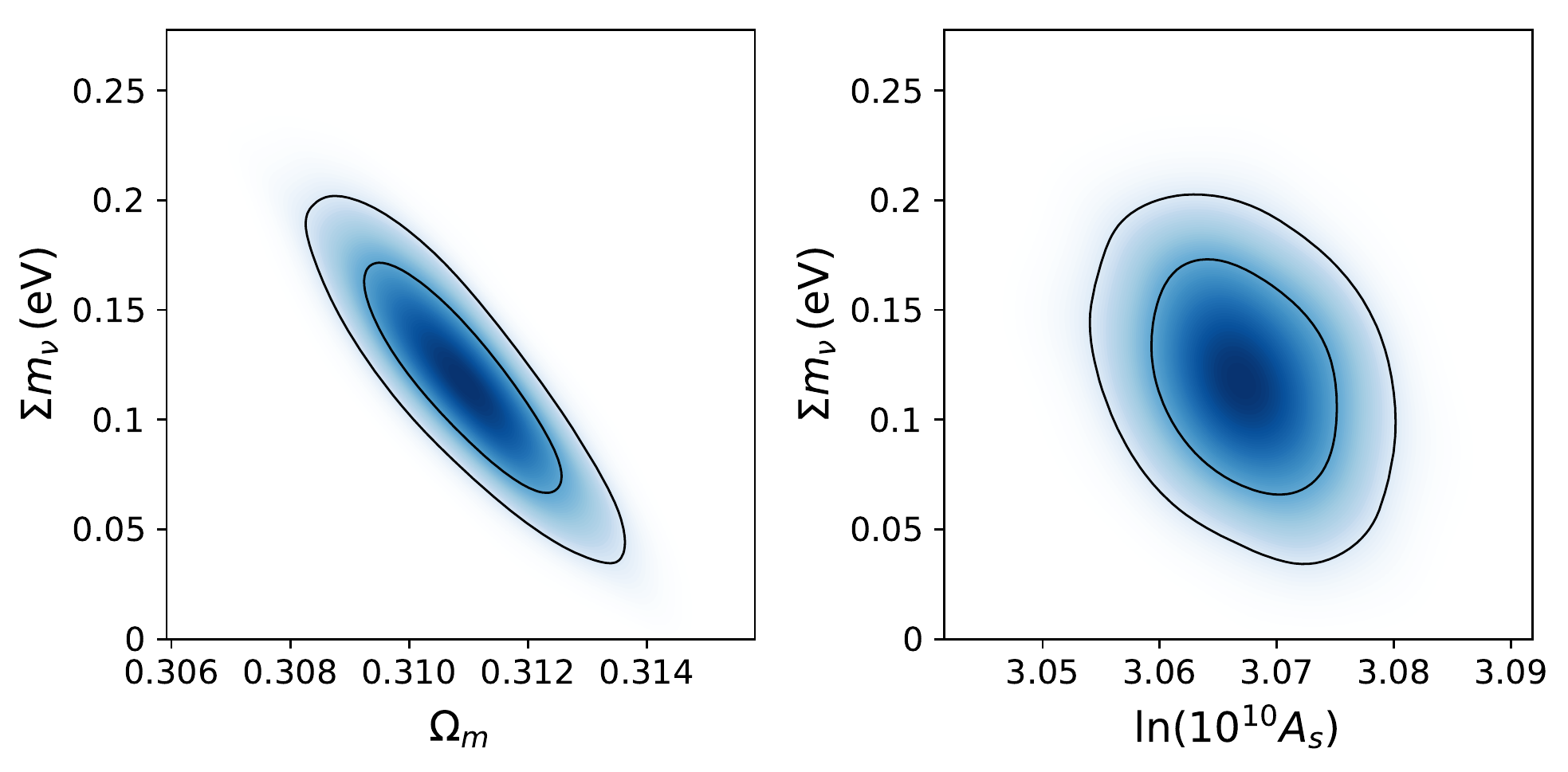}
\caption{Left: the 1 ans 2$\sigma$ level contours of $\Sigma m_\nu$ versus $\Omega_m$, marginalising over 	the other relevant cosmological parameters, from our joint MCMC analysis. Right: same as left, but with $\Sigma m_\nu$ versus $\ln(10^{10} A_s)$ instead. The inferred 
neutrino mass is $m_{\nu} = 0.119 \pm 0.034$ eV at 67\% confidence level.}
\label{MCMC}
\end{figure*}

Finally, in order to make our predictions more robust, we have explored the relatively sensitive dependence of neutrino mass on $\Omega_m$ by modifying CosmoMC \cite{Lewis:2002ah, Lewis:2013hha} to include our measured correlation length data points with their errors shown in Figure 2. The data we used are the ratios between $r_0$ in a universe with and without neutrinos. In the case without neutrinos, $r_0$ is calculated assuming independent values of $A_s$ (where $A_s$ refers to the amplitude of the primordial power-spectrum) and $\Omega_ch^2$ than the case with massive neutrinos. This naturally increases the error bars compared with the case with similar values of $A_s$ and $\Omega_ch^2$ for the numerator and denominator of $r_0$ ratio. The likelihood is assumed to be Gaussian. We also use the \textit{Planck} CMB measurements including low-$\ell$ and high-$\ell$ temperature and polarization data. For convenience, the mass of the neutrinos is assumed to be entirely one species. Flat priors are used for $\Sigma m_\nu$, $\Omega_ch^2$, and $\ln(10^{10}A_s)$. The other cosmological parameters are fixed: $\Omega_bh^2=0.02280$,
$\tau=0.066$, $n_s=0.9667$,
$H_0=67.73\,\mathrm{km\,s^{-1}\,\rm{Mpc}^{-1}}$. The Gelman and Rubin statistics for convergence $R-1$ is 0.01138.
Figure \ref{MCMC} presents the results of this analysis. Here we present 1$\sigma$ and 2$\sigma$ level contours of the neutrino mass wrt $\Omega_m$. The inferred neutrino mass is $\Sigma m_{\nu} = 0.119 \pm 0.034$ eV at 67\% confidence level. 
	
At face value, our result is in best agreement with the inverted based hierarchy minimum mass of $\simeq$$0.1eV$ based on neutrino oscillation work, and in some tension with the minimum value for the normal hierarchy of $\sim$$0.056eV$ at the $95\%$ confidence level. If ongoing laboratory results from the NOvA and T2K collaborations continue to favor a normal neutrino mass hierarchy \cite{Adamson:2017qqn,Abe:2016fic}, however, then our result may imply a somewhat higher neutrino mass than minimum oscillation based value of $m_{\nu} \simeq 0.056eV$. Alternatively, the standard relic density underestimates the total cosmological neutrino density by a factor of two, implying an additional light neutrino contribution. This possibility, however, is not supported by measurements of the CMB anisotropy that strictly imply only three relic neutrino species \cite{Planck Collaboration:2019} as in the Standard Model of particle physics.
	
We aim to improve upon our result by jointly fitting the cluster correlation function enhancement and the cluster abundance suppression.  We will also use the simulations to make an assessment of the preferred mass range for constraining relic neutrinos, as the halo mass function is so steep (proportional to $M^{-4}$) in the cluster regime that selecting groups $>10^{13}M_\odot$ may prove more fruitful than expanding the survey volume. An elegant group definition for the SDSS has been devised by Zhao et al \cite{Zhao:2015ecx} for which the mean bias is $b\simeq 2$, similar to the luminous red galaxy (LRG) mass scale, so that we may use these large samples with redshift measurements for a significant improvement.

We can also improve upon our precision for $\sum m_{\nu}$ by defining a more accurate weak lensing based mass-richness relation for a representative subset of our complete cluster sample.  Such work will be possible with the upcoming wide field J-PAS survey \cite{Benitez:2014ibt}, which can go beyond the careful SDSS based work of Simet et al (2017) to greater depth and higher angular resolution over the Northern sky. 
	
We can see that pursuing the above practical improvements is really well motivated given how close we are already to achieving the accuracy required to definitively distinguish between the inverted and normal hierarchies and the fully mass degenerate minimum of $\sum m_{\nu} \simeq 0.15eV$. Furthermore, with future data reaching higher redshift we may also examine the redshift dependence of the combined correlation amplitude and cluster abundance evolution to test whether the lightest relic neutrino eigenstate remains relativistic until today.

\section*{Acknowledgments}
We are grateful for useful conversations with Neta Bahcall, Andrew Cohen,  Daniel J. Eisenstein, Lars Hernquist, Lam Hui, Francisco-Shu Kitaura, Marilena Loverde, Airam Marcos-Caballero, Abraham Loeb, Enrique Martinez, Alberto Rubino, Martin Schmaltz, David Spergel, Henry Tye, Jun Qing Xia and Francisco Villaescusa. We also thank two anonymous referees for their insightful comments.
The work of R.E. was previously supported by Hong Kong University through the CRF Grants of the Government of the Hong Kong SAR under HKUST4/CRF/13 and is currently supported by the Institute for Theory and Computation at the Center for Astrophysics at Harvard University. 
GFS acknowledges the IAS at HKUST and the Laboratoire APC-PCCP, Universit\'{e} Paris Diderot and Sorbonne Paris Cit\'{e} (DXCACHEXGS) and also the financial support of the UnivEarthS Labex program at Sorbonne Paris Cit\'{e} (ANR-10-LABX-0023 and ANR-11-IDEX-0005-02). TJB thanks IAS fr generous hospitality.

\end{document}